\documentclass[twocolumn,aps,pra,showpacs,preprintnumbers,bibnotes]{revtex4-1}

\usepackage[T1]{fontenc}

\usepackage{natbib}
\usepackage{graphicx}
\usepackage{bm}
\usepackage{color}
\usepackage{amsmath}

\DeclareMathOperator{\re}{Re}
\DeclareMathOperator{\im}{Im}

\newcommand\unit[2]{\ensuremath{#1~\mathrm{{#2}}}}

\newcommand\Ket[1]{\ensuremath{|{#1}\rangle}}

\newcommand\Isotope[2]{\ensuremath{^{#1}\mathrm{#2}}}
\newcommand\Li{\Isotope{6}{Li}}
\newcommand\K{\Isotope{40}{K}}
\newcommand\Rb{\Isotope{87}{Rb}}

\renewcommand{\vec}[1]{\ensuremath{\bm{#1}}}

\newcommand\NA{\ensuremath{\mathrm{NA}}}
\newcommand{\osim}{\ensuremath{\mathord{\sim}}}

\newcommand\TrapFreq{\ensuremath{\nu_t}}
\newcommand\RecoilEnergy{\ensuremath{E_\mathrm{rec}}}
\newcommand\RecoilFrequency{\ensuremath{\nu_\mathrm{rec}}}

\begin{document}

\title{Low-noise optical lattices for ultracold $^6$Li}

\author{S. Blatt}
\email{sblatt@physics.harvard.edu}
\author{A. Mazurenko}
\author{M. F. Parsons}
\author{C. S. Chiu}
\author{F. Huber}
\altaffiliation{Present address: IPG Photonics Corp., Oxford, MA 01540}
\author{M. Greiner}
\affiliation{
  Department of Physics, Harvard University,
  Cambridge, Massachusetts, 02138, USA}

\date{\today}

\begin{abstract}
  We demonstrate stable, long-term trapping of fermionic $^6$Li atoms in an optical lattice with MHz trap frequencies for use in a quantum gas microscope. Adiabatic release from the optical lattice in the object plane of a high-numerical-aperture imaging system allows a measurement of the population distribution among the lowest three bands in both radial directions with atom numbers as low as $7\times 10^2$. We measure exponential ground band heating rates as low as 0.014(1)~$\mathrm{s^{-1}}$ corresponding to a radial ground state $1/e$ lifetime of $71(5)~\mathrm{s}$, fundamentally limited by scattering of lattice photons. For all lattice depths above 2 recoil, we find radial ground state lifetimes $\ge 1.6 \times 10^6$ recoil times.
\end{abstract}

\maketitle

Trapped quantum particles are used widely in modern atomic physics from quantum information science~\cite{leibfried03,wineland13} and quantum simulations of many-body physics~\cite{bloch08,blatt12} to atomic clocks~\cite{ludlow15} and studies of fundamental physics~\cite{gabrielse12,loh13}.
All of these experiments benefit from long motional coherence times, often because they enable coherent rather than statistical averaging of results.
Such long times require preparation of the particles in a well-defined motional state of the trap, ideally by ground state cooling.
The traps must not only be stable enough to prevent the particles from escaping, but they
should preserve the carefully prepared state of motion for as long as possible.

Its light mass, $m$, makes fermionic \Li{} particularly suited to quantum simulations in optical lattices~\cite{bloch12}.
All energy scales in an optical lattice are naturally parametrized by the lattice recoil energy, $h\RecoilFrequency{}$, and recoil frequency, $\RecoilFrequency{} = h / (8 m a^2)$, associated with the geometric lattice spacing, $a$, where $h$ is Planck's constant.
For the same tunneling rate in recoil units, the absolute tunneling rate is a factor of 14.5 (6.7) faster for \Li{} than for \Rb{} (\K{}) atoms.
Assuming typical atomic lifetimes of one minute, it will thus be possible to study thermalization processes and superexchange dynamics on timescales much longer than previously accessible~\cite{esslinger10}.
Here, we demonstrate an intensity-stable, high-power optical lattice for \Li{} atoms.
The optical lattice is designed for a quantum gas microscope where individual sites of the optical lattice and individual atoms can be resolved in fluorescence microscopy~\cite{bakr09, sherson10}.

Fluorescence imaging of \Li{} with resonant light at $\lambda_p = \unit{671}{nm}$ is hampered by the large resonant recoil energy $E_p = h^2 / (2 m \lambda_p^2) = h \times \unit{74}{kHz}$.
Each scattering event adds $\ge 2 E_p$ on average, regardless of the atom's motional state~\cite{wolf00}.
This recoil heating makes it challenging to keep the atoms cold enough to suppress tunneling while scattering $\mathcal{O}(10^4)$ resonant photons to form an image.
For this reason, we have to combine a laser cooling scheme with deep optical lattices and MHz trap frequencies.
Trap frequencies in the MHz regime require a trapping laser with low intensity noise
because parametric heating rates due to laser intensity fluctuations increase quadratically with trap frequency~\cite{gehm98,gehm00}.
Trap quality can be degraded further through thermal lensing effects in the lattice optics at high intensities.
To address these challenges, we implemented a high-power and low-noise lattice laser system based on Yb-doped fiber amplifiers seeded with an intensity-stable Nd:YAG laser.

In this paper, we use this laser system to demonstrate stable trapping of the two-dimensional ground band in the object plane of our quantum gas microscope with $1/e$ lifetimes exceeding one minute.
The corresponding heating rates are measured with a sensitive band mapping technique~\cite{greiner01,esslinger10}.
The high numerical aperture (\NA{} = 0.87) of our imaging system allows measurement of the band populations for total atom numbers as low as $7\times 10^2$.
We find that the measured heating rates are consistent with a rate equation model based on measured intensity-noise spectra and spontaneous scattering rates, which dominate the heating at high trap depths.

\begin{figure}
  \centering
  \includegraphics[width=\columnwidth]{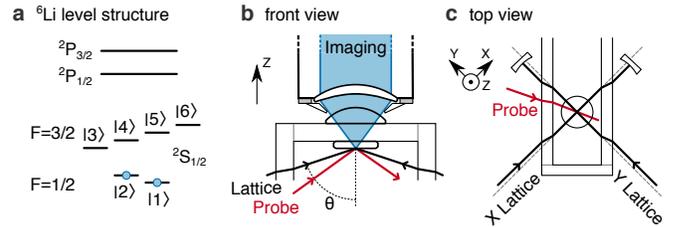}
  \caption{(color online). (a) Simplified \Li{} level structure (not to scale). We use an incoherent mixture of the two lowest states in the \Li{} $^{2}\mathrm{S}_{1/2}$ ground state manifold. (b) Front view of the glass cell. The \unit{1064}{nm} optical lattices reflect off of a superpolished substrate at shallow angle of incidence ($70^\circ$). Fluorescence induced by the probe beam is collected through a high \NA{} imaging system whose object plane is $\osim\unit{10}{\mu m}$ below the substrate. (c) Top view of the glass cell and definition of the lab frame $(X,Y,Z)$. The optical lattices are retroreflected along $X$ and $Y$.}
  \label{fig:setup}
\end{figure}

The lowest two hyperfine manifolds in \Li{} are shown in Fig.~\ref{fig:setup}(a), and the states are commonly labeled \Ket{1}-\Ket{6} according to their energy splitting in a magnetic bias field.
In our experiment, we load an incoherent mixture of atoms in states \Ket{1} and \Ket{2} into a high-power optical dipole trap (\unit{1064}{nm}, \unit{300}{W}) and transfer its focus into the center of the fused silica vacuum cell shown in Fig.~\ref{fig:setup}(b) and (c).
The atoms are then transferred into a crossed optical dipole trap formed by incoherent light derived from a superluminescent diode at \unit{780}{nm}, whose short coherence length avoids fringing when passing through the imaging optics~\cite{huber14}.
The crossed dipole trap is located \unit{80}{\mu m} below the object plane of a high-resolution microscope system and the sample is evaporated further in this trap at a magnetic field of \unit{300}{Gauss}.
We then load a single layer of an optical accordion lattice~\cite{huber14} and decrease the accordion incidence angle from $88^\circ$ to $70^\circ$.
The angular change simultaneously compresses the sample and transports it to a distance of \unit{10}{\mu m} from the superpolished mirror that is the final lens of the imaging system.
We then adiabatically load $\osim 3.5\times 10^3$ atoms into a three-dimensional optical lattice along directions $X$, $Y$, and $Z$.
As shown in Fig.~\ref{fig:setup}(c), the $\lambda = \unit{1064}{nm}$ optical lattice beams reflect off of the superpolished mirror resulting in a standing wave along the axial direction with spacing $a_z = \frac{\lambda}{2 \cos{\theta}} = \unit{1.56}{\mu m}$ for an incidence angle $\theta = 70^\circ$.
By retroreflecting each lattice beam, we obtain a non-interfering optical lattice along the $X$ and $Y$ axes with equal spacings $a_x = a_y = \frac{\lambda}{2\sin\theta} = \unit{569}{nm}$.
The radial (axial) lattice spacing corresponds to a recoil frequency $\RecoilFrequency{} = \unit{25.9}{kHz}$ (\unit{3.4}{kHz}).

\begin{figure}
  \centering
  \includegraphics[width=\columnwidth]{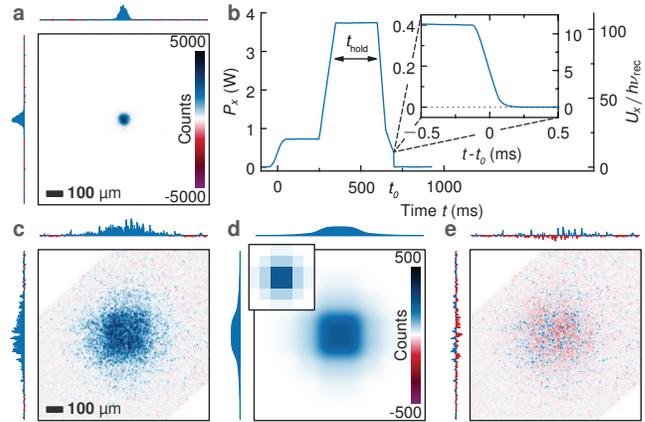}
  \caption{(color online). Starting with $3.5\times 10^3$ atoms in the ($X$, $Y$) plane shown in the in-trap fluorescence image (a, dark image subtracted), we apply the adiabatic ramp shown in panel (b) and obtain the band-mapping image (c). These images are fit (d) with a convolution of in-trap distribution and band map (the deconvolved band map is shown in the inset). The fit residuals are shown in panel (e). All images represent the same region, and images (c)-(e) share the same color bar. We show cuts through the center of each image in the margins.}
  \label{fig:bandmap}
\end{figure}

To image the atoms \emph{in situ}, we apply the probe beam on the D$_2$ transition -- containing two frequencies resonant with both hyperfine manifolds in the ground state -- as indicated in Fig.~\ref{fig:setup}(b).
The fluorescence is collected on an intensified CCD camera to produce the image in Fig.~\ref{fig:bandmap}(a).
Note that we have increased the field of view of our 0.87 \NA{} infinite conjugate ratio imaging system at the expense of resolution by demagnifying the image.

We load the optical lattice by ramping up the lattice powers $P_x$ and $P_y$ to $\unit{0.8}{W}$ adiabatically, as shown in Fig~\ref{fig:bandmap}(b).
Each lattice beam's power is controlled by two independent servos, and control over the lattice power is handed over automatically depending on the setpoint.
For setpoints below $P_\mathrm{ex} = \unit{0.95}{W}$, we use a low-power high-bandwidth servo. For setpoints above $P_\mathrm{ex}$, we use a high-power low-bandwidth servo.
After loading the lattice, $P_x$ and $P_y$ are changed adiabatically to a holding power and the atoms are held in the corresponding lattice for a variable time, $t_\mathrm{hold}$, during which they are heated by lattice intensity fluctuations and spontaneous scattering of lattice photons.
At the end of the experiment, we release the atoms from the lattice using a ramp that is adiabatic with respect to the band gaps of the lattice, but fast compared to the residual harmonic confinement timescales.
The high-bandwidth servo allows releasing the atoms within \unit{200}{\mu s}, using the ramp shown in the inset of Fig.~\ref{fig:bandmap}(b).
At the end of the ramp, the atoms are allowed to expand ballistically for \unit{1.7}{ms}, after which we apply a short probe pulse to obtain the image in Fig.~\ref{fig:bandmap}(c).

This image shows clear rectangular features due to the band edges of the radial lattice that are convolved with the in-trap density distribution.
We model the band map distribution by flat rectangular features with widths set by the radial lattice spacings, convolved with a two-dimensional Gaussian distribution.
The resulting fit and its residuals are shown in panels (d) and (e), and cuts through the center of each image are shown in the margins.
We extract the radial band populations $C_{ij}$ from the fit amplitude in the corresponding Brillouin zone, shown in the inset of panel (d).

\begin{figure}
  \centering
  \includegraphics[width=\columnwidth]{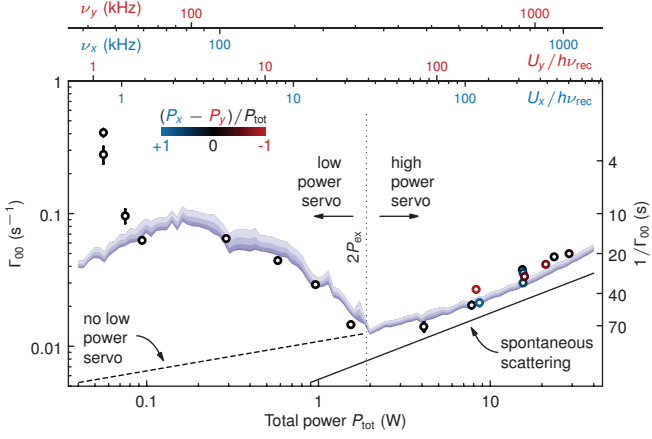}
  \caption{(color online). Ground band heating rates $\Gamma_{00}$ as a function of total lattice power $P_\mathrm{tot} = P_x + P_y$. The marker color indicates the fractional lattice power mismatch $(P_x - P_y) / P_\mathrm{tot}$. The heating rates are obtained by fitting exponential decays to the ground band populations $C_{00}$ from band mapping images after holding atoms in the lattice for a variable time. The error bars indicate the statistical uncertainty from the fits. We calibrate the lattice trap frequency $\nu_x$ ($\nu_y$) via lattice modulation spectroscopy for different lattice powers $P_x$ ($P_y$), leading to the calibrated scales shown at the top for $P_x = P_y$. The shaded areas indicate the results from a rate equation model based on the measured intensity-noise spectra for different angular mismatch $\Delta\theta$. The dashed line indicates the simulated heating rate without intensity control. The solid line shows the asymptotic contribution from spontaneous scattering of lattice light $\propto \sqrt{P_\mathrm{tot}}$.}
  \label{fig:heating}
\end{figure}

By varying the hold time in the lattice and observing how the radial ground state population $C_{00}$ varies with time, we can measure an exponential heating rate.
The resulting heating rates for different lattice powers are shown in Fig.~\ref{fig:heating}.
We find exceptionally low heating rates and correspondingly long lifetimes of the radial ground band of up to \unit{71(5)}{s}.
Even at the deepest trap depths and MHz trap frequencies required to implement single-site resolved imaging in a quantum gas microscope, the radial ground band still has a lifetime of \unit{20}{s}.
For the deepest trap depths the heating process is dominated by exponential heating due to spontaneous scattering of lattice photons (solid line in Fig.~\ref{fig:heating}).

Because the axial optical lattice is formed by two lattice beams, the position of the atoms is sensitive to intensity fluctuations on either lattice beam.
The spacing of the axial lattice formed by each beam depends on the angle of incidence $\theta$ as argued above.
If the angles of incidence for $X$ and $Y$ are mismatched by $\Delta\theta \equiv |\theta_x - \theta_y|$, uncorrelated intensity noise on $P_x$ and $P_y$ will lead to fluctuations of the axial trap minimum, causing fast heating that has the same dependence on motional quantum number as spontaneous photon scattering.
By including all of these processes in a rate equation model based on measured intensity-noise spectra and a geometric estimate of $\Delta\theta \le 1.2^\circ$, we obtain the shaded region in Fig.~\ref{fig:heating},
whose lower boundary corresponds to $\Delta\theta = 0$.
In the model, the radial ground state population $C_{00}$ is calculated by integrating over the axial state populations of the three-dimensional trap (see Appendix~\ref{sec:heating-rate-model}).

The model also assumes a trap depth along the radial and axial axes which we calculate from measured trap frequencies under the assumption of sinusoidal modulation.
The trap frequencies are calibrated via lattice modulation spectroscopy~\cite{huber14}.
The modulation spectra also show cross-modulation peaks at the base frequency and confirm the effect.
For lattice powers corresponding to trap depths below $2 h\RecoilFrequency{}$, the
model deviates from the experimental data because the harmonic oscillator approximation inherent in the rate equation model does not describe shallow lattices well.

\begin{figure}
  \centering
  \includegraphics[width=\columnwidth]{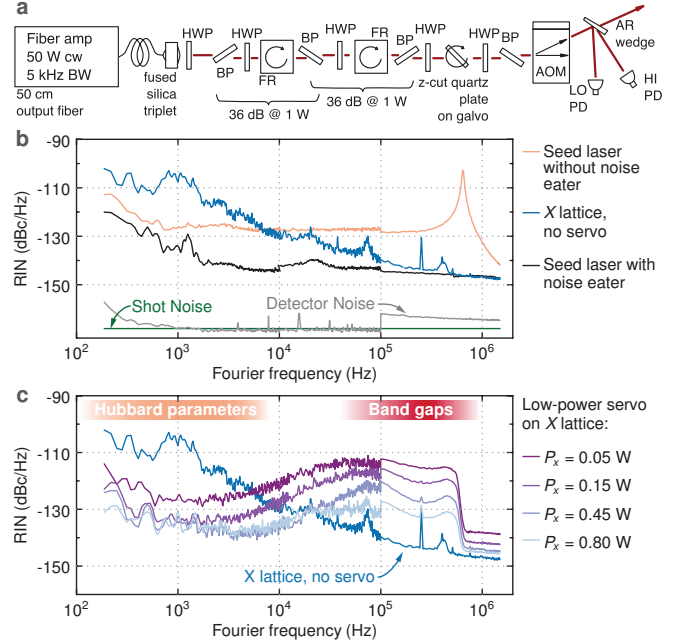}
  \caption{(color online). (a) Lattice optics setup including fiber amplifier, collimation optics, optical isolators, and power actuators as described in the main text. (b) Relative intensity-noise (RIN) spectra for the $X$ lattice laser (the $Y$ laser has similar features) and the seed laser taken as described in Appendix~\ref{sec:rin-measurements}. (c) For low lattice powers, the high-bandwidth analog servo suppresses intensity noise below \unit{10}{kHz}. The servo is conservatively tuned to work across 1.5 orders of magnitude in setpoint. The servo gain and bandwidth decrease with decreasing setpoint, leading to a larger contribution of noise from the servo electronics. The increased noise at frequencies corresponding to the band gaps of the optical lattice leads to the higher heating rates for low lattice powers in Fig.~\ref{fig:heating}. At the same time, the servo reduces the noise below the seed laser RIN in the frequency range corresponding to the Hubbard model parameters~\cite{pichler12}.}
  \label{fig:rin}
\end{figure}

The lattice beams are derived from Yb-doped fiber amplifiers (Nufern), seeded by an intensity-stable Nd:YAG non-planar ring oscillator laser (Innolight Mephisto).
Because of the high optical power requirements, the beam path schematically shown in Fig.~\ref{fig:rin}(a) is designed to be mechanically and thermally stable.
At high intensities, many materials show thermal lensing effects that result in beam pointing and focusing changes on fast timescales.
For these reasons, the beam paths (as well as the vacuum chamber) use fused silica optics which are less susceptible to thermal lensing because of the material's high thermal conductivity, low thermal expansion coefficient, and low index of refraction sensitivity to temperature.

Thermal effects in $\mathrm{Tb}_3\mathrm{Ga}_5\mathrm{O}_{12}$ (TGG) optical isolators result in a loss of isolation at high optical powers~\cite{dooley12} which can lead to damage to the fiber amplifiers from backreflections.
To provide sufficient isolation, we use two stages of TGG Faraday rotators (FR) with fused silica Brewster polarizers (BP) and $\lambda/2$ waveplates (HWP).
To suppress thermal variation in the isolators, the full optical power is always incident on the isolators.

The optical power in the lattice beams is controlled in two stages.
As a first stage, we use a low-bandwidth, high-dynamic-range actuator built from a z-cut birefringent quartz plate attached to a galvo motor.
The quartz plate acts as a variable waveplate (Berek compensator) and in combination with a polarizer becomes a variable attenuator with a dynamic range of $\osim\unit{20}{dB}$ over $\osim 4.5^\circ$ of rotation.

For the second stage of intensity stabilization, we use a large-active-area acousto-optical modulator (AOM, Crystal Technology 3080-197) as the actuator.
Thermal effects in the $\mathrm{TeO}_2$ AOM crystal can also result in pointing drifts on slow time scales.
Since the optical lattices are focused to \unit{80}{\mu m} waists at the position of the atoms, such pointing noise would cause position fluctuations of the trap center.
To ameliorate these effects, the lattice beams are focused through the AOM crystal by adjusting the distance between the fiber tip and the air-spaced fused silica collimator (Optosigma 027-0510).

We detect the optical power on two independent, shot-noise-limited transimpedance amplifiers using \unit{1064}{nm} enhanced InGaAs photodiodes (PD, Hamamatsu G8370-01) with gain just small enough to cover bandwidths up to \unit{700}{kHz}.
The detected photovoltages are then used as the input to two independent servo loops.

For optical powers above $P_\mathrm{ex} = \unit{0.95}{W}$, we do not require fast control over the lattice depth but require that the laser intensity is as passively stable as possible.
In this regime, we feed back on the angular rotation of the quartz plate using a slow digital feedback loop.
The low-bandwidth (small-signal \unit{3}{dB} point $\osim\unit{2}{kHz}$) actuator ensures that we cannot write noise onto the laser at frequencies comparable to the trap frequencies.

In Fig.~\ref{fig:rin}(b), we show laser intensity power spectral densities for the $X$ lattice and seed lasers.
The spectra are normalized to the DC power and are commonly referred to as relative intensity-noise (RIN) spectra (see Appendix~\ref{sec:rin-measurements}).
The prominent relaxation oscillation peak in the seed laser spectrum becomes well-suppressed when engaging the built-in intensity-noise eater.
After seeding the fiber amplifiers, the intensity stability is degraded by \unit{10-20}{dB} for Fourier frequencies below \unit{1}{MHz}.
Noise spikes from the switching power supplies driving the fiber amplifier pump diodes can be strongly suppressed by low-pass filtering the power supplies (MPE DS26387).
Acoustic pointing noise on the fiber tip from power-supply fans is suppressed by removing the power supplies from the amplifier enclosure.

For optical powers below $P_\mathrm{ex}$, we use analog feedback on the rf amplitude driving the AOM with an rf mixer and limit the servo's bandwidth by inserting a steep low-pass filter at \unit{500}{kHz} (LPF-B0R5).
The bandwidth limitation ensures that no electronic noise gets written onto the lattice amplitude at frequencies in the MHz regime.
For small power setpoints, the AOM servo has a nonlinear transfer function due to the use of a frequency mixer.
This nonlinearity results in setpoint-dependent gain and reduced bandwidth for small setpoints, which is partially compensated by a Schottky-diode-based linearization circuit in the controller.
By conservatively tuning the loop, we are able to control the lattice depth over 2.5 orders of magnitude at the expense of slightly degraded noise spectra and lifetime for small lattice depths, as seen in Fig.~\ref{fig:heating}.
Here, such a large dynamic range is useful to obtain the adiabatic band mapping images. Additionally, the low-power servo suppresses noise at frequencies relevant for quantum simulation of Hubbard models~\cite{pichler12}.
If required, the servo tuning can be reoptimized for even longer lifetimes for lattice depths of interest.

In conclusion, we have demonstrated stable trapping of \Li{} atoms in the radial ground band of \unit{1064}{nm} optical lattices spanning 2.5 orders of magnitude in trap depth.
We measure ground band populations with sensitivity down to $7 \times 10^2$ atoms and the $1/e$ lifetime $\tau$ in the ground band can exceed one minute for deep lattices and is longer than \unit{10}{s} ($2\pi\RecoilFrequency{}\tau > 1.6\times 10^6$) for all lattice depths above $2 h\RecoilFrequency{}$.
These heating rates are one to two orders of magnitude smaller than in ion traps with comparable trap frequencies~\cite{brownnutt14}, with smaller distances to the nearest surface (\unit{10}{\mu m} here).
In a three-dimensional optical lattice with MHz trap frequencies, we demonstrate radial ground band lifetimes that are comparable with the longest trap lifetimes measured in optical dipole traps~\cite{ohara99} and one-dimensional optical lattices~\cite{gibbons08}.
The heating rates are well-explained by a rate equation model based on the measured intensity-noise spectra.
These spectra can be further tailored by servo design for application in quantum simulation experiments with \Li{}.
Optical traps with MHz frequencies enable ion-trap-like spectral addressability~\cite{wineland79,leibfried03}, are compatible with proximity to surfaces, and may have applications in achieving strong coupling to high-quality-factor mechanical resonators~\cite{camerer11}.

We thank W. Setiawan and K. Wooley-Brown for early contributions, G. Jotzu for discussions, and B. Lincoln of Nufern for his hospitality and openness.
We acknowledge support by ARO DARPA OLE, ARO MURI, and NSF. A.M., M.F.P., and C.S.C. are supported by NSF GRFP, and S.B. acknowledges support by the Harvard Quantum Optics Center.

\appendix

\section{RIN Measurements}
\label{sec:rin-measurements}

To measure the RIN, we used a shot-noise limited transimpedance amplifier design~\cite{scott01} based on an InGaAs photodiode (Hamamatsu G8370-01), a fast operational amplifier with \unit{800}{MHz} gain-bandwidth product (OPA843), and a transimpedance gain of \unit{330}{\Omega}. To reduce thermal drift from interference between the photodiode and its cover glass, we removed the photodiode window. Residual light contamination was attenuated by placing an interference filter (Semrock FF01-1020/LP-25, ND = 5 for visible light) before the photodiode.

We measured RIN power spectral densities (PSDs) by putting the output of such a photodetector onto a Fourier transform precision voltmeter (SR760, varying resolution band width [RBW] \unit{65}{Hz} below \unit{12.5}{kHz}, and \unit{500}{Hz} above) below \unit{100}{kHz} or a battery powered RF spectrum analyzer (Anritsu MS2721A, RBW = \unit{10}{Hz}) above \unit{100}{kHz} Fourier frequency. The noise spectra were normalized to a frequency bin width of \unit{1}{Hz} (assuming white noise in each bin), and the SR760 spectrum was converted from $\mathrm{dBV_{rms}/\sqrt{Hz}}$ to $\mathrm{dBm/Hz}$, to be comparable with the spectrum analyzer. Such spectra were then further normalized to the optical carrier power using the DC voltage measured with an RMS voltmeter.
For RIN measurements, we typically apply \unit{12}{mW} of optical power to the photodiode (resulting in \unit{2.0}{V} DC signal) to reduce the shot noise level below the noise floor $P_\text{floor}$ of the RF spectrum analyzer (typically $P_\text{floor} \simeq \unit{-155}{dBm}$ at RBW = \unit{10}{Hz}). Battery powered operation reduces noise from external sources and makes the measurements compatible with the SR760 results, allowing us to combine data from both devices in the plots shown in Fig.~\ref{fig:rin}.

\section{Heating Rate Model}
\label{sec:heating-rate-model}

From the measured RIN spectra, we calculate single-particle heating rates in deep optical lattices due to trap intensity fluctuations and cross-modulation for mismatched trap centers~\cite{gehm98,ohara99,gehm00,supplemental}.
We estimate heating rates from spontaneous scattering of lattice photons for optical lattices in the Lamb-Dicke regime from standard expressions for laser cooling~\cite{cohen-tannoudji92,wolf00,leibfried03,supplemental}.

We then combine all heating rates in a three-dimensional rate equation for the probabilities $P_{\vec{n}}$ to occupy the motional state $\vec{n} = (n_x, n_y, n_z)$ of the form
\begin{equation}
  \label{eq:20}
  \dot{P}_{\vec{n}}(t) = \sum_{\Delta\vec{n}} [R_{\vec{n} \leftarrow \vec{n} + \Delta\vec{n}} - R_{\vec{n} + \Delta\vec{n} \leftarrow \vec{n}}] P_{\vec{n}}(t).
\end{equation}

To compare our measured band mapping data against the heating rate coefficients, we numerically solve Eqn.~(\ref{eq:20}).
We sum $P_{\vec{n}}$ over the vertical direction to get horizontal band populations $C_{ij}$ which we can then directly compare to the band map fit coefficients.

Most of the heating rate coefficients depend on the RIN PSD. For the low power servo settings, we linearly interpolate (in linear units) between measured spectra such as the ones in Fig.~\ref{fig:rin}(c).
For the high power servo settings, we use the power-independent RIN spectra from Fig.~\ref{fig:rin}(b).
The PSDs are linearly interpolated (in linear units) at the Fourier frequency of interest.
For this interpolation, the Fourier frequencies of interest are multiples of the trap frequencies, $(\nu_x, \nu_y, \nu_z)$.
The trap frequencies are calibrated against the measured optical power via lattice modulation spectroscopy.

The initial distribution among trap states is assumed to be Boltzmann with a temperature adjusted to match the measured band map pictures at short times.
We limit the state space to states with energies below the lowest modulation depth (along the $y$ direction, estimated from $\nu_y$) and assume that an atom is completely lost once it is heated to states with higher quantum numbers.

The three-dimensional rate equation is propagated from the initial condition
and the vertically averaged populations $C_{ij}$ are fit with exponential loss curves.
The resulting exponential decay rates are then compared to the experimental data in Fig.~\ref{fig:heating}.

The parameter with the largest uncertainty is the mismatch between the angles of incidence of the two lattice beams $\Delta\theta$ responsible for the cross-modulation heating contribution.
With the simulation, we generate heating rates for several values below a conservative upper limit $\Delta\theta \le 1.2^\circ$, leading to the shaded areas in Fig.~\ref{fig:heating}.

\clearpage
\pagebreak
\widetext
\begin{center}
\textbf{\large Supplemental Material for:\\
Low-noise optical lattices for ultracold \Li{}} \\
\vspace{1em}
S. Blatt,$^*$ A. Mazurenko, M. F. Parsons, C. S. Chiu, F. Huber,$^\dag$ and M. Greiner \\
\textit{Department of Physics, Harvard University,
Cambridge, Massachusetts, 02138, USA} \\
(Dated: \today)
\end{center}
\vspace{1cm}

\setcounter{section}{0}
\setcounter{equation}{0}
\setcounter{figure}{0}
\setcounter{table}{0}
\makeatletter
\renewcommand{\theequation}{S\arabic{equation}}
\renewcommand{\thefigure}{S\arabic{figure}}
\renewcommand{\bibnumfmt}[1]{[S#1]}
\renewcommand{\citenumfont}[1]{S#1}
\makeatother

\twocolumngrid

\section*{Lattice heating mechanisms}
\label{sec:latt-heat-mech}

We present a short summary of results derived in Refs.~\cite{Ssavard97,Sgehm98,Sgehm00} and extend them to derive an expression for cross-modulation of overlapped but non-interfering optical lattices with incommensurable periodicity in the tight-binding limit. To ensure we account for all numerical factors, we present all equations with frequencies $\nu$ in units of Hz in the sense of $\cos(2\pi\nu t)$. Rates and rate coefficients $\Gamma$ still carry factors of $2\pi$ in the sense of $\exp(-\Gamma t)$ and are presented in units of $s^{-1}$.

\subsection{Noise-induced heating in harmonic traps}
\label{sec:heat-harm-traps}

Intensity fluctuations of the dipole trapping laser will result in fluctuations of the trap frequency and produce parametric transitions between states of the same parity. In the limit of small fluctuations in a tight-binding optical lattice we can approximate the potential for an atom with mass $M$ around a lattice minimum $x_0$ as harmonic with trap frequency \TrapFreq{}. We model the intensity fluctuations with a small fluctuation $\epsilon$ and write
\begin{equation}
  \label{eq:3}
  V(x) = \frac{M}{2} (2\pi\TrapFreq)^2 (1 + \epsilon) (x-x_0)^2
\end{equation}
In this limit, we can also confine ourselves to transitions between harmonic trap states $n$ and $n\pm 2$. The rate coefficients $R_{n\pm2 \leftarrow n}$ for these transitions can be calculated in time-dependent perturbation theory as~\cite{Sgehm98}
\begin{equation}
  \label{eq:1}
  \begin{aligned}
  R_{n\pm 2 \leftarrow n} &= \frac{\pi^2}{8} \TrapFreq^2 S_\epsilon(2\TrapFreq) (n+1 \pm 1)(n\pm 1), \\
  & \equiv R_\epsilon(\TrapFreq) (n+1 \pm 1)(n\pm 1).
  \end{aligned}
\end{equation}
where $S_\epsilon(2\TrapFreq)$ is the relative power spectral density of the laser intensity in (linear) units of $10^{-\mathrm{dBc}/10} / \mathrm{Hz}$ evaluated at twice the trap frequency. For an infinitely deep trap, these rate coefficients result in an exponential increase in mean energy with the rate
\begin{equation}
  \label{eq:2}
  \Gamma_\epsilon \equiv \langle \dot{E} \rangle / \langle E \rangle = \pi^2 \TrapFreq^2 S_\epsilon(2\TrapFreq).
\end{equation}

If the harmonic trap center $x_0$ is subject to fluctuations $\delta$, we write
\begin{equation}
  \label{eq:4}
  V(x) = \frac{M}{2} (2\pi\TrapFreq)^2 [x - (x_0 + \delta)]^2,
\end{equation}
which induces transitions between trap states of opposite parity. The largest rate coefficients are given by
\begin{equation}
  \label{eq:5}
  \begin{aligned}
  R_{n\pm1 \leftarrow n} & = 4 \pi^4 \frac{M}{h} \TrapFreq^3 S_\delta(\TrapFreq) (n+\frac{1}{2}\pm\frac{1}{2}) \\
  & = \pi^2 \TrapFreq^2 S_{\delta / a}(\TrapFreq) (n + \frac{1}{2}\pm\frac{1}{2}), \\
  & \equiv R_\delta(\TrapFreq) (n + \frac{1}{2}\pm\frac{1}{2}),
  \end{aligned}
\end{equation}
to show that we can normalize the position noise by  the harmonic oscillator length $a^2 \equiv \hbar / (M 2\pi\TrapFreq)$ to get a more intuitive result. Assuming an infinitely deep trap again, we find a linear increase in mean energy with rate
\begin{equation}
  \label{eq:6}
  \begin{aligned}
  \langle \dot{E} \rangle / (h \TrapFreq) &= (h\TrapFreq)^{-1} \times 4\pi^4 M \TrapFreq^4 S_\delta(\TrapFreq) \\
  &= \pi^2 \TrapFreq^3 S_{\delta/a}(\TrapFreq).
  \end{aligned}
\end{equation}

\subsection{Cross-modulation}
\label{sec:cross-modulation}

Consider the case of two harmonic traps with different trap centers and different trap frequencies. The total potential is then again harmonic with
\begin{equation}
  \label{eq:7}
  \begin{aligned}
  V(x) = &\frac{M}{2} \left[( 2\pi\nu_1)^2 (x - x_1)^2 +
    ( 2\pi\nu_2)^2 (x - x_2)^2 \right] \\
  \equiv & \frac{M}{2} (2\pi)^2 (\nu_1^2 + \nu_2^2) \\
  & \times \left[(x - x_0)^2 + r(1-r) (x_1^2 + x_2^2)\right], \\
  \end{aligned}
\end{equation}
where we have defined the trap frequency asymmetry parameter $r = \nu_1^2 / (\nu_1^2 + \nu_2^2)$ and the new trap center $x_0 = r x_1 + (1-r) x_2$. If we allow intensity fluctuations $\epsilon_i$ in each trap frequency as in Eqn.~\eqref{eq:3}, we see that the trap center $x_0$ also fluctuates and find (to first order in $\epsilon_i$) relative trap frequency and trap center fluctuations with
\begin{equation}
  \label{eq:8}
  \begin{aligned}
  \epsilon & = r\epsilon_1 + (1-r)\epsilon_2, \\
  \delta & = r(1-r)(\epsilon_1 - \epsilon_2) (x_1 - x_2). \\
  \end{aligned}
\end{equation}

Clearly, the trap center will only fluctuate if both traps have non-zero frequency, if their centers are not matched and only if the intensity fluctuations are uncorrelated. Assuming independent noise processes we find power spectral densities
\begin{equation}
  \label{eq:10}
  \begin{aligned}
    S_\epsilon(2 \TrapFreq) &= r^2 S_{\epsilon_1}(2\TrapFreq) + (1-r)^2 S_{\epsilon_2}(2\TrapFreq), \\
    S_\delta(\TrapFreq) & = [r(1-r) \Delta x]^2 [S_{\epsilon_1}(\TrapFreq) + S_{\epsilon_2}(\TrapFreq)], \\
  \end{aligned}
\end{equation}
with trap mismatch $\Delta x \equiv x_1 - x_2$.

\subsection{Application to surface-reflected lattices}
\label{sec:appl-surf-refl}

We are now ready to apply the heating rates derived in the previous Sections to the special case of our surface-reflected optical lattices. We model the site-local potential for a deep optical lattice (in the tight-binding regime) as a three-dimensional harmonic oscillator. Experimentally, we can directly measure determine the ground to first excited band energy difference using parametric heating measurements as described in the main text. The intensity-noise induced heating rates along the horizontal lattice directions $\hat{x}$ and $\hat{y}$ are
\begin{equation}
  \label{eq:11}
  \begin{aligned}
    R^x_\epsilon(\nu_x) &= \frac{\pi^2}{8} \nu_x^2 S_\epsilon^x(2\nu_x), \\
    R^y_\epsilon(\nu_y) &= \frac{\pi^2}{8} \nu_y^2 S_\epsilon^y(2\nu_y), \\
  \end{aligned}
\end{equation}
where we have neglected the small contribution of the $x$ ($y$) lattice beam to the other trap frequency $\nu_y$ ($\nu_x$).

Along the vertical direction, the surface lattice beams also form an optical lattice with spacing given by the laser wavelength $\lambda$ and the angles $\theta_x$ and $\theta_y$ with respect to the surface.
The position of the $n$-th lattice minimum away from the surface is given by $z_n(\theta_i) = \lambda / (2 \sin\theta_i)$.
If we assume a small angle mismatch $\theta_x = \theta - \Delta\theta/2$ and $\theta_y = \theta + \Delta\theta/2$, we find that the minima produced by the lattice beams differ by $\Delta z = z_n(\theta_x) - z_n(\theta_y) \simeq z_n(\theta) \Delta\theta / \tan\theta$.
Using the arguments presented in Sec.~\ref{sec:cross-modulation}, we can immediately see that in addition to the intensity noise heating, the position of the $n$-th vertical lattice minimum will shake as well.
We find heating rates for the vertical direction as
\begin{equation}
  \label{eq:12}
  \begin{aligned}
    R^z_\epsilon(\nu_z) =& \frac{\pi^2}{8} \nu_z^2 [r^2 S_\epsilon^x(2\nu_z) + (1-r)^2 S_\epsilon^y(2\nu_z)] \\
    R^z_\delta(\nu_z) =& 4 \pi^4 \frac{M}{h} \nu_z^3 \left[r(1-r) \frac{\Delta\theta}{\tan\theta} z_n(\theta)\right]^2 \\
    & \times [S_\epsilon^x(\nu_z) + S_\epsilon^y(\nu_z)].
  \end{aligned}
\end{equation}
From Eqn.~\eqref{eq:12}, we can immediately see that the cross-modulation of the vertical lattice center can be problematic because it scales with the cube of the vertical trap frequency and is proportional to the intensity noise power spectral density evaluated at the fundamental of the trap frequency.
However, careful angle-matching of the incident lattice beams should be able to reduce this heating rate dramatically.

\subsection{Single-photon scattering}
\label{sec:single-phot-scatt}

The optical dipole trap will also produce heating from lattice photon scattering. Each scattering event will produce an average \emph{total} energy increase of $r_h \RecoilEnergy$, where $r_h = 2$ for an optically thin atomic cloud~\cite{Swineland79,Scohen-tannoudji92,Swolf00}.
In a motional state resolved picture in the Lamb-Dicke regime, the scattering events will produce heating that is described well by transitions between states $n\pm 1\leftarrow n$ to first order in the Lamb-Dicke parameter squared $\eta^2 = \RecoilEnergy / (h \TrapFreq)$.
This treatment leads to transition rate coefficients \cite{Sstenholm86,Sleibfried03}
\begin{equation}
  \label{eq:13}
  \begin{aligned}
    R_{n\pm 1\leftarrow n} =& \frac{1}{3}\frac{r_h}{2} \eta^2 [\Gamma_\text{sc}(\Delta) + \Gamma_\text{sc}(\Delta \mp 2\pi\TrapFreq)] \\
    & \times \left(n+\frac{1}{2}\pm\frac{1}{2}\right),\\
     \equiv & R_\mathrm{sc} \left(n+\frac{1}{2}\pm\frac{1}{2}\right)
  \end{aligned}
\end{equation}
with identical dependence on $n$ as the rate coefficient for trap center fluctuations.
Here, the scattering rate $\Gamma_\text{sc}(\Delta)$ is to be understood as the steady-state scattering rate solution from the optical Bloch equations for detuning $\Delta$ from atomic resonance.
For the far-detuned case, $\Gamma_\text{sc}(\Delta \mp 2\pi\TrapFreq) \simeq \Gamma_\text{sc}(\Delta) \equiv \Gamma_\mathrm{sc}$.
Since we do not distinguish between scattering events due to individual lasers, we added a factor of $1/3$ in Eqn.~\eqref{eq:13}, which assumes isotropic heating and ensures that the total energy increases as~\cite{Sgehm98}
\begin{equation}
  \label{eq:18}
  \begin{aligned}
  \langle \dot{E} \rangle & = \sum_j \langle \dot{E}_j \rangle = \sum_j \sum_{n=0}^\infty h \nu_j R^j_\mathrm{sc} P_{n_j}(t) = r_h \RecoilEnergy \Gamma_\mathrm{sc},
  \end{aligned}
\end{equation}
where we used that the populations are normalized to $\sum_{n_j} P_{n_j} = 1$.

To relate the scattering rate $\Gamma_\text{sc}$ to experimentally accessible observables, we note that the scattering rate is proportional to the overall optical dipole trap depth $U$.
The atomic polarizability $\alpha$ for an alkali atom ground state is well described by a single Lorentz oscillator model~\cite{Sgrimm99} with resonant frequency $\omega_0$ and linewidth $\Gamma$:
\begin{equation}
  \label{eq:15}
  \alpha = 6\pi\epsilon_0 c^3 \frac{\Gamma / \omega_0^2}{\omega_0^2 - \omega^2 - i(\omega^3/\omega_0^2) \Gamma}.
\end{equation}
The proportionality coefficient between $\Gamma_\text{sc}$ and $U$ is given by the ratio of the real and imaginary parts of $\alpha$~\cite{Sgrimm99}, and we find for \Li{} ground state atoms in a $\lambda = 2\pi c/\omega = \unit{1064}{nm}$ optical dipole trap that
\begin{equation}
  \label{eq:16}
  \xi \equiv \frac{\hbar \Gamma_\text{sc}}{U} = \frac{2 \im\alpha}{\re\alpha} \simeq 1.09 \times 10^{-8}.
\end{equation}

In summary, the heating rate for lattice photon scattering in the Lamb-Dicke regime is
\begin{equation}
  \label{eq:17}
  R^j_\mathrm{sc} = \frac{r_h}{3} \xi \frac{U}{\hbar} \frac{\RecoilEnergy}{h \nu_j}
\end{equation}

Our best estimate for the total dipole trap depth $U$ is the lattice modulation depth along the vertical direction which for a sinusoidal modulation is related to the measured vertical trap frequency by
\begin{equation}
  \label{eq:19}
  \frac{U}{\hbar} = 2\pi \nu^z_\mathrm{rec} \left(\frac{\nu_z}{2\nu^z_\mathrm{rec}}\right)^2,
\end{equation}
where $\nu^z_\mathrm{rec} = h/(8 M d_z^2)$ is the geometric recoil frequency related to the vertical lattice spacing $d_z = \lambda / (2\sin\theta)$.

\end{document}